\documentclass[a4paper,12pt]{article}
\usepackage{geometry}

\usepackage{graphicx,empheq,authblk,color} 
\usepackage[hidelinks]{hyperref}

\usepackage[nottoc]{tocbibind}

\title{Reduced dimensional description of hydromagnetic turbulence capturing higher order fluid moments}
\author{Arindam Saikia$^\dagger$}
\author{Rupak Mukherjee$^\ddagger$}
\affil{Department of Physics, Sikkim University, Gangtok, Sikkim, India}

\date{\today}

\begin{document}

\maketitle

\begin{abstract}
In this paper, Chandrasekhar's deductive theory of turbulence is extended to the case of hydrodynamics and hydromagnetics, with higher order fluid moments. We include the contributions from correlation tensors at two different points of space and two different points in time to account for the energy transfer between Fourier modes with different $k$ values. We start with three governing equations using the first three moments of Vlasov equation, which gives us particle, momentum and energy conservation relations, respectively. We obtain the scalar representations of correlation tensors for hydrodynamics as well as hydromagnetics. Finally we pick an example from literature where the energy spectra alone was not capable of explaining the observations reported by the authors.
\end{abstract}

\vspace{3mm}

\vspace{1mm}

\noindent
{\bf {$^\dagger$email:}} arindamsaikia16@gmail.com\\
{\bf {$^\ddagger$email:}} rmukherjee@cus.ac.in

\newpage

\tableofcontents



\section{Introduction}

Articulating turbulence mathematically is a challenging topic due to the complexity in deriving precise equations that encapsulate the chaotic nature of fluid flows involved. Kolmogorov's theory provides a heuristic model to understand turbulence that is able to capture the energy cascade process across different scales. However a rigorous phenomenological theory remains necessary, as Kolmogorov’s model provides statistical insight rather than a deterministic picture that fully describe turbulent motions.\\

An alternative methodology for modeling such systems was pioneered by Subrahmanyan Chandrasekhar, who took a more systematic and mathematically rigorous approach, advancing our understanding especially in magneto-hydrodynamics. His approach involved formulating equations solely containing correlation scalars. His basic idea was to introduce correlations in the velocity components ($u_i$) at two different points ($x$ and $x^\prime$) and at two different times ($t$ and $t^\prime$) and deduce a single equation through his framework to describe the turbulent dynamics \cite{chandrasekhar:1956}. He also showed that at $t-t^\prime=0$, i.e., when the time difference in taken to be zero, the differential equations take the form one supposes on Kolmogorov's similarity principles. In a later communication, Chandrasekhar extended his theory for hydromagnetics as well, following the same methodology \cite{chandrasekhar:1955a, chandrasekhar:1955b}. \\

One can consider Vlasov equation and take moments of different order. The zeroth order moment of Vlasov equation gives continuity equation. The first order moment gives momentum equation and so on. We know, each moment equation contains a higher order moment of the distribution function. For example, the continuity equation (zeroth order moment of Vlasov equation) contains density (zeroth order moment of the distribution function) as well as mean velocity (first order moment of the distribution function). Similarly the momentum equation (first order moment of Vlasov equation) contains density (zeroth order moment of the distribution function), mean velocity (first order moment of the distribution function) and pressure (second order moment of the distribution function), and so on. Thus the set of moment equations of Vlasov equation is never closed, and is known as the BBGKY hierarchy problem. To make the set of fluid equations well-posed, we often take resort to a closure equation, that is to express a higher order moment of the distribution function, in terms of the lower order moments. For example, under certain circumstances, pressure (second order moment of the distribution function) can be written in terms of density (zeroth order moment of the distribution function). It is known that the set of moment equations involving higher order moments describe the properties of Vlasov equation better than a set of moment equations involving lower order moments \cite{wang:2015, cagas:2017, wang:2018, dong:2019, ng:2020, struchtrup:2019, struchtrup:2022}. For example, the set of continuity, momentum and pressure equations describe a fluid medium better than a set of continuity and momentum equation only.\\

Our primary motivation for this work was to extend Chandrasekhar's work for higher order moments of the Vlasov equation. An example case for working with lower order moments would be the closure at the first order using Fick's law forms \cite{hammett:1990}. Taking higher order moments has been shown to take us closer to a kinetic picture of the system rather than a purely statistical fluid picture which, as has been shown in the later sections, fails to account for and predict a couple of expected behaviours from complex hydromagnetic systems. In our work, we have derived them for continuity equation, momentum conservation equation and energy conservation equation with a Hammett-Perkins type closure \cite{hammett:1990}. A further extension of the work in terms of including the third moment is in progress. In addition, following Chandrasekhar's work on hydromagnetics, we extend our study with higher order fluid moments for magnetised fluid as well.


\section{Chandrasekhar's Theory of Hydrodynamics}

Chandrasekhar argued that characterizing turbulence solely through its energy spectrum, $F(k)$, is insufficient for a complete description, as it neglects the critical phase relationships between different Fourier components of the velocity field. These phase relationships play an essential role in the transfer of energy between modes. To address this gap, he introduced correlations between velocity components measured at two distinct spatial points and times, thereby incorporating these phase dependencies into the analysis. His theory was formulated under the assumption of isotropy and considered scenarios where a stationary state exists, offering a more comprehensive framework for understanding turbulent dynamics.

Under the assumption that the correlations depend, apart from the vector $\xi ( = r'' - r')$ joining the two points, only on the difference in the times, $|t'' - t'|$, the following correlation tensors were introduced : 

\begin{eqnarray}
\label{eq:equation1}
&& Q_{ij} = \overline{u_i(r',t')u_j(r'',t'')}\\
\label{eq:equation2}
&& T_{ij;k} = \overline{u_i(r',t')u_j(r',t')u_k(r'',t'')}\\
\label{eq:equation3}
&& P_{ij} = \overline{\overline{\omega}(r',t')u_i(r'',t'')u_j(r'',t'')}\\
\label{eq:equation4}
&& Q_{ij;kl} = \overline{u_i(r',t')u_j(r',t')u_k(r'',t'')u_l(r'',t'')}
\end{eqnarray}
where $\overline{\omega} = p/\rho$ and  $u_i$ denote the components of turbulent velocity. Here, the semi-colon notation on the tensor subscripts serves as a convention for separating the indices measured at $t'$ or $t''$ coordinates. In the later case, the indices are kept on the right side of the semi-colon. This notation doesn't represent a covariant derivative or contraction in this use case.

In addition to these correlation tensors, an additional statistical hypothesis had to be put forward in order to reduce the final relation into a single equation in terms of $Q$ alone (here $Q$ is the scalar representation of the tensor $Q_{ij}$) : 

\begin{equation}
\label{eq:equation5}
Q_{ij;kl} = Q_{ik}Q_{jl} + Q_{il}Q_{jk} + Q_{ij}(0,0)Q_{kl}(0,0)
\end{equation}

This hypothesis states that the quadruple moment $Q_{ij;kl}$ is related to the second-order moment $Q_{ij}$ as in a \textit{normal distribution.}

Now we are equipped to tackle the equation of motion for the fluid under study: 

\begin{equation}
\label{eq:equation6}
\frac{\partial u_i'}{\partial t'} + \frac{\partial}{\partial x_k'}u_i'u_k' = - \frac{\partial \overline{\omega}'}{\partial x_i'} + \nu \nabla^2u_i' 
\end{equation}

The above equation expressed in terms of correlation tensors takes the form: 

\begin{equation}
\label{eq:equation7}
\pm \frac{\partial Q_{ij}}{\partial t} = \frac{\partial}{\partial \xi_k}T_{ik;j} + \nu \nabla^2_\xi Q_{ij}
\end{equation}

Now, using scalar representations for these tensor equations [1], the final equation is obtained,

\begin{equation}
\label{eq:equation8}
\frac{\partial}{\partial r}(\frac{\partial^2}{\partial t^2} - \nu^2D^2_5)Q = -2Q\frac{\partial}{\partial r}D_5Q.
\end{equation}

Here, $D_n$ is the Laplacian Operator for a spherically symmetric function in an $n$-dimensional Euclidean space. Hence, $D_5^2$ represents the square of the Laplacian operator in a 5-dimensional Euclidean space.


\section{Chandrasekhar's Theory of Hydromagnetics}

A similar approach is applied to the case of hydromagnetics, analogous to Chandrasekhar’s earlier work in hydrodynamics, but this time beginning with the following set of equations: 

\begin{align}
\label{eq:equation9}
\frac{\partial u_i}{\partial t} + \frac{\partial}{\partial x_k}\left(u_iu_k - h_ih_k\right) &= -\frac{\partial \overline{\omega}}{\partial x_i} + \nu \nabla^2 u_i \\
\text{and,} \quad \frac{\partial h_i}{\partial t} + \frac{\partial}{\partial x_k}\left(h_iu_k - u_ih_k\right) &= \lambda \nabla^2 h_i
\end{align}

again, $u_i$ denotes the components of turbulent velocity, $h_i$ the components of the magnetic field divided by $(4\pi\rho/\mu)^{\frac{1}{2}}$, $\rho$ the density, $P$ the pressure and $\mu$,$\nu$ and $\sigma$ are the coefficients of magnetic permeability, kinematic viscosity and electrical conductivity, respectively, and 

\begin{align}
\label{eq:equation10}
\lambda = 1/4\pi\mu\sigma \hspace{0.4cm} \text{and}  \hspace{0.4cm} \overline{\omega} = P/\rho + \frac{1}{2}|h|^2
\end{align}

In addition, the following divergence conditions are used

\begin{equation}
    \label{eq:equation11}
    \frac{\partial u_i}{\partial x_i} = 0 \hspace{0.4cm} \text{and} {\hspace{0.5cm}} \frac{\partial h_i}{\partial x_i} = 0
\end{equation}

This time around, the amount of correlation tensors required to represent the system are as follows: 

\begin{empheq}[right=\empheqrbrace]{align}
& 
\begin{aligned}
& Q_{ij} = \overline{u_i(r',t')u_j(r'',t'')} \\
& H_{ij} = \overline{h_i(r',t')h_j(r'',t'')} \\
& P_{ij} = \overline{\overline{\omega}(r',t')u_i(r'',t'')u_j(r'',t'')} \\
& \Pi_{ij} = \overline{\overline{\omega}(r',t')h_i(r'',t'')h_j(r'',t'')} \\
& T_{ij;k} = \overline{u_i(r',t')u_j(r',t')u_k(r'',t'')} \\
& S_{ij;k} = \overline{h_i(r',t')h_j(r',t')u_k(r'',t'')} \\
& F_{ij;k} = \overline{\left[h_i(r',t')u_j(r',t') - u_i(r',t')h_j(r',t')\right]h_k(r'',t'')} \\
& Q_{ij;kl} = \overline{u_i(r',t')u_j(r',t')u_k(r'',t'')u_l(r'',t'')} \\
& H_{ij;kl} = \overline{h_i(r',t')h_j(r',t')h_k(r'',t'')h_l(r'',t'')} \\
& \hspace{-1cm} R_{ij;kl} = \overline{\left[h_i(r',t')u_j(r',t') - u_i(r',t')h_j(r',t')\right]\left[h_k(r'',t'')u_l(r'',t'') - u_k(r'',t'')h_l(r'',t'')\right]}
\end{aligned}
\end{empheq}

which allows us to reduce the dynamics into a set of two scalar equations: 

\begin{eqnarray}
\label{eq:equation14}
&&\frac{\partial}{\partial r} \left(\frac{\partial^2}{\partial t^2} - \nu^2D_5^2\right)Q = -2Q \frac{\partial}{\partial r }D_5Q - 2H \frac{\partial}{\partial r}D_5H \\
\label{eq:equation15}
&& \left(\frac{\partial^2}{\partial t^2} - \lambda^2D_5^2\right)H = -2QD_5H - 2HD_5Q -2\frac{\partial Q}{\partial r}\frac{\partial H}{\partial r}
\end{eqnarray}

Equations \ref{eq:equation14} and \ref{eq:equation15} are the required equations governing $Q$ and $H$.


\section{Basic Ideas of the Proposed Theory}

In line with Chandrasekhar's theory, we introduce correlation tensors to account for phase coupling, connecting measurements taken at two separate spatial points and two distinct moments in time. This inclusion directly reflects the presence of phase coupling between these points.

To do so, we start with the Vlasov equation, which describes the simplest possible case of linear one-dimensional electrostatic waves given by : 

\begin{equation}
\label{eq:equation16}
\frac{\partial f}{\partial t} + v \frac{\partial f}{\partial x} + \frac{e}{m} E \frac{\partial f}{\partial x} = 0
\end{equation}

Proceeding with \ref{eq:equation16}, we take moments of said equation, and arrive at the following three relations \cite{hammett:1990}: 

\begin{eqnarray}
\label{eq:equation17}
&& \frac{\partial n}{\partial t} + \frac{\partial}{\partial x}(un) = 0\\
\label{eq:equation18}
&& \frac{\partial}{\partial t} (m n u) + \frac{\partial}{\partial u} (u m n u) = -\frac{\partial P}{\partial x} + e n E + \frac{\partial S}{\partial x}\\
\label{eq:equation19}
&& \frac{\partial P}{\partial t}  + \frac{\partial}{\partial x} (u P) = -(\Gamma - 1) (P+S) \frac{\partial u}{\partial x} - \frac{\partial q}{\partial x}
\end{eqnarray}

Equations \ref{eq:equation17}, \ref{eq:equation18}, and \ref{eq:equation19} represent particle, momentum, and energy conservation relations, respectively. The heat flux moment '$q$' is considered as prescribed, effectively closing this set of coupled equations.

Here, $q = m \int dv \, f(v-u)^3$, and the dissipative momentum flux $S$ is given by $S = mn\mu \frac{\partial u}{\partial x}$ 

In the following sections, hydrodynamics and hydromagnetics will be extended to include these moments while using a correlation tensor approach to represent the resulting dynamics.


\section{Extension of hydrodynamics for higher order moments}

Before we proceed with the derivations, we will define the following correlation tensors: 

\begin{minipage}{0.45\textwidth}
\begin{empheq}[right=\empheqrbrace]{align}
& 
\begin{aligned}
& C_j = \overline{mn'u_j''} \nonumber \\
& F_{ik} = \overline{u_i'u_k'} \\
& N_{ij} = \overline{P_i'u_j''} \\
& \Sigma_{ij} = \overline{enR_i'u_j''} \\
& D_{jl} = \overline{mn'u_j''u_l''} 
\end{aligned}
\end{empheq}
\end{minipage}
\begin{minipage}{0.45\textwidth}
\begin{empheq}[right=\empheqrbrace]{align}
& 
\begin{aligned}
& M_{jl;i} = \overline{P_i'u_j''u_l''} \\
& \tilde{\Sigma}_{jl;i} = \overline{enE_j'u_j''u_l''} \\
& \widetilde{M}_{ij} = \overline{u_i'P_i'u_j''} \\
& X_{ij} = \overline{q_i'u_j''} \\
& M^0_{jl;i} = \overline{u_i'P_i'u_j''u_l''}
\end{aligned}
\end{empheq}
\end{minipage}

Starting with \ref{eq:equation17}, the continuity equation, and introducing the following sign conventions to describe measurements at two different points: ${'}$ representing measurements at spatial point $r^{'}$ and at time $t^{'}$, whereas, ${''}$ represents $r^{''}$ and $t^{''}$ points, we expand the derivative term and arrive at:

\begin{equation}
\label{eq:equation1}
\frac{\partial n'}{\partial t'} + u_i' \frac{\partial}{\partial x_i'} n' + n' \frac{\partial}{\partial x_i'} u_i' = 0
\end{equation}

Multiplying the equation bu $u_j''$ and averaging all the terms across ensembles , transforms the operators as below ( with $\xi = (r''-r'$ )) and the $\pm$ sign depending on whether $t'' > t'$ or $t'' < t'$ :  

\begin{equation}
\label{eq:equation1}
\pm \frac{\partial}{\partial t} C_j - \langle u \rangle \frac{\partial}{\partial \xi_i} C_j - \langle \rho \rangle \frac{\partial}{\partial \xi_i} Q_{ij} = 0
\end{equation}

Now, utilizing the fact that a scalar representation of the above tensor equation is possible in the case of homogeneous isotropic turbulence, where the tensors are invariant under arbitrary rotations and reflections of the reference axes \cite{chandrasekhar:1953}, we obtain the following expression:

\begin{equation}
\label{eq:equation1}
\pm \frac{\partial C}{\partial t} - \langle \rho \rangle \frac{\partial}{\partial r} Q = 0
\end{equation}

Here (in eq.22), the second term goes to zero due to the implication of equation of continuity for an incompressible fluid.

Again, multiplying equation(21) with $u_j''u_l''$ and averaging all the terms across ensembles, we get

\begin{equation}
\label{eq:equation1}
\pm \frac{\partial}{\partial t} D_{jl} - \langle u \rangle \frac{\partial}{\partial \xi_i} D_{jl} + \langle \rho \rangle \frac{\partial}{\partial \xi_i} T_{jl;i} = 0
\end{equation}

We obtain the following scalar equation from equation (24)

\begin{equation}
\label{eq:equation1}
\pm \frac{\partial}{\partial t} D - \langle u \rangle \frac{\partial}{\partial r} D + \langle \rho \rangle \left( r \frac{\partial}{\partial r} + 5 \right) T = 0
\end{equation}

where, we have used the general form of a solenoidal isotropic tensor of second order \cite{chandrasekhar:1953}, in order to find the corresponding scalar operators for $D_{jl}$.

Next, Using definition of $S$ in (18) and expanding the derivative terms, we get the following form

\begin{equation}
\label{eq:equation1}
m n' \frac{\partial u_i'}{\partial t} + u_i' \frac{\partial (m n')}{\partial t} + m n' \frac{\partial (u_i' u_k')}{\partial x_k} + u_i' u_k' \frac{\partial (m n')}{\partial x_k'} = -\frac{\partial P_i'}{\partial x_i'} + e n E_i' + m n' \mu \frac{\partial^2 u_i'}{\partial x_i'^2}
\end{equation}

As before, we multiply with $u_j''$ and take ensemble average,

\begin{equation}
\label{eq:equation1}
\pm \langle \rho \rangle \frac{\partial}{\partial t} Q_{ij} \pm \langle u \rangle \frac{\partial}{\partial t} C_j - \langle \rho \rangle \frac{\partial}{\partial \xi_k} T_{ik;j} - F_{ik} \frac{\partial}{\partial \xi_k} C_j = \frac{\partial}{\partial \xi_i} N_{ij} + \Sigma_{ij} + \mu \langle \rho \rangle \frac{\partial^2}{\partial \xi_i^2} Q_{ij}
\end{equation}

In scalar representation, taking $C_j$ and $N_{ij}$ terms to be equal to zero from equation of continuity,

\begin{equation}
\label{eq:equation1}
\pm \langle \rho \rangle \frac{\partial}{\partial t} Q \pm \langle u \rangle \frac{\partial}{\partial t} C - \langle \rho \rangle (r \frac{\partial}{\partial r} + 5) T  = \Sigma + \mu \langle \rho \rangle D_7 Q
\end{equation}

Similarly , multiplying with $u_j''u_l''$ and averaging, we get,

\begin{equation}
\label{eq:equation1}
\pm \langle \rho \rangle \frac{\partial}{\partial t} (-T_{ij;l}) \pm \langle u \rangle \frac{\partial}{\partial t} D_{jl} - \langle \rho \rangle \frac{\partial}{\partial \xi_k} Q_{ik;jl} - F_{ik} \frac{\partial}{\partial \xi_k} D_{jl} = \frac{\partial}{\partial \xi_i} M_{jl;i} + \widetilde{\Sigma}_{jl;i} + \mu \langle \rho \rangle \frac{\partial^2}{\partial \xi_i^2} (-T_{jl;i})
\end{equation}

And in scalar representation,

\begin{equation}
\label{eq:equation1}
\mp \langle \rho \rangle \frac{\partial}{\partial t} T \mp \langle u \rangle \frac{\partial}{\partial t} D - \langle \rho \rangle \hat{O_3} Q - F \cdot \frac{\partial}{\partial r} D = (r \frac{\partial}{\partial r} + 5) M + \widetilde{\Sigma} - \mu \langle \rho \rangle D_7 T
\end{equation}

lastly, we perform multiplication with $u_j''$ on equation (19) in the form :

\begin{equation}
\label{eq:equation1}
\frac{\partial P_i'}{\partial t'} + \frac{\partial}{\partial x_i'} (u_i' P_i') = -\Gamma P_i' \frac{\partial u_i'}{\partial x_i'} - \Gamma m n' \mu \frac{\partial^2 u_i'}{\partial x_i'^2} + P_i' \frac{\partial u_i'}{\partial x_i'} + m n' \mu \frac{\partial^2 u_i'}{\partial x_i'^2} - \frac{\partial q_i'}{\partial x_i'}
\end{equation}

To get, 

\begin{equation}
\label{eq:equation1}
\pm \frac{\partial}{\partial t} [N_{ij}] - \frac{\partial}{\partial \xi_i} \widetilde{M}_{ij} = \Gamma \langle P \rangle \frac{\partial}{\partial \xi_i} Q_{ij} - \Gamma \langle \rho \rangle \mu \frac{\partial^2}{\partial \xi_i^2} Q_{ij} - \langle P \rangle \frac{\partial}{\partial \xi_i} Q_{ij} + \langle \rho \rangle \mu \frac{\partial^2}{\partial \xi_i^2} Q_{ij} + \frac{\partial}{\partial \xi_i} X_{ij}
\end{equation}

Which gives us,
\begin{equation}
\label{eq:equation1}
\pm \frac{\partial N}{\partial t} = \frac{\partial}{\partial r} (\widetilde{M} + \Gamma \langle P \rangle Q - \langle P \rangle Q + X ) - (1-\Gamma)\langle \rho \rangle \mu D_5 Q
\end{equation}

Multiplying equation (26) with $u_j''u_l''$ and averaging

\begin{eqnarray}
\label{eq:equation1}
\pm \frac{\partial}{\partial t} M_{jl;i} - \frac{\partial}{\partial \xi_i} M_{jl;i}^0 = \Gamma \langle P \rangle \frac{\partial}{\partial \xi_i} (-T_{jl;i}) - \Gamma \langle \rho \rangle \mu \frac{\partial^2}{\partial \xi_i^2} (-T_{jl;i}) - \langle P \rangle \frac{\partial}{\partial \xi_i} (-T_{jl;i}) \nonumber \\
+ \langle \rho \rangle \mu \frac{\partial^2}{\partial \xi_i^2} (-T_{jl;i}) + \frac{\partial}{\partial \xi_i} \tilde{X}_{jl;i}
\end{eqnarray}

Giving us,

\begin{equation}
\label{eq:equation1}
\pm \frac{\partial M}{\partial t} = \left(r \frac{\partial}{\partial r} + 5 \right) \left(M^0 + (1 - \Gamma) \langle P \rangle T - \tilde{X}\right) + \left(\Gamma D_7 - D_5\right) \langle \rho \rangle \mu T
\end{equation}

We can now simplify equations (23), (25), (28), (30), (33), (35) to get three concise equations that govern the dynamics,

\begin{equation}
\label{eq:equation1}
\pm \langle \rho \rangle \frac{\partial Q}{\partial t} + \langle u \rangle \langle \rho \rangle \frac{\partial}{\partial r} Q - \langle \rho \rangle \left( r \frac{\partial}{\partial r} + 5 \right) T - F \cdot \frac{\partial}{\partial r} C = \frac{\partial}{\partial r} N + \Sigma + \mu \langle \rho \rangle D_5 Q
\end{equation}

\[
\left( - \frac{\partial^2}{\partial t^2} \pm \mu \frac{\partial}{\partial t} D_7 \pm \langle u \rangle \left( r \frac{\partial}{\partial r} + 5 \right) \langle \rho \rangle \frac{\partial T}{\partial t} \pm (\langle u \rangle^2 + F) \frac{\partial}{\partial t} (\frac{\partial}{\partial r} D) \right) = 
\]
\begin{equation}
\label{eq:equation1}
\left( r \frac{\partial}{\partial r} + 5 \right)^2 \left( M^0 + (1 - \Gamma) \langle \rho \rangle T - \tilde{X} \right) + \left( r \frac{\partial}{\partial r} + 5 \right) (\Gamma D_7 - D_5) \langle \rho \rangle \mu T \pm \frac{\partial \tilde{\Sigma}}{\partial t} \pm \langle \rho \rangle \frac{\partial}{\partial t} \hat{O}_3 Q
\end{equation}

\begin{equation}
\label{eq:equation1}
\pm \frac{\partial N}{\partial t} = \frac{\partial}{\partial r} \left( \widetilde{M} + \Gamma \langle P \rangle Q - \langle P \rangle Q + X \right) - (1 - \Gamma) \langle \rho \rangle \mu D_5 Q
\end{equation}


\section{Extension of hydromagnetics for higher order moments}

With a foundational framework for hydrodynamics established using equations (36), (37), and (38), and Chandrasekhar’s relations for hydromagnetics presented in Section III through equations (14) and (15), we can now propose a more refined model for hydromagnetic turbulence by integrating these equations. This approach enables us to derive a higher-order representation, providing a more detailed depiction of Chandrasekhar’s hydromagnetic theory : 

\begin{equation}
    \label{eq:equation1}
    (\frac{\partial^2}{\partial t^2} - \lambda^2D_5^2)H = -2QD_5H - 2HD_5Q -2\frac{\partial Q}{\partial r}\frac{\partial H}{\partial r}
\end{equation}

\begin{equation}
\label{eq:equation1}
\pm \langle \rho \rangle \frac{\partial Q}{\partial t} + \langle u \rangle \langle \rho \rangle \frac{\partial}{\partial r} Q - \langle \rho \rangle \left( r \frac{\partial}{\partial r} + 5 \right) (T+S) - F \cdot \frac{\partial}{\partial r} C = \frac{\partial}{\partial r} N + \Sigma + \mu \langle \rho \rangle D_5 Q
\end{equation}

\[
\left( - \frac{\partial^2}{\partial t^2} \pm \mu \frac{\partial}{\partial t} D_7 \pm \langle u \rangle \left( r \frac{\partial}{\partial r} + 5 \right) \langle \rho \rangle \frac{\partial T}{\partial t} \pm (\langle u \rangle^2 + F) \frac{\partial}{\partial t} \frac{\partial}{\partial r} D \right) = 
\]
\begin{equation}
\label{eq:equation1}
\left( r \frac{\partial}{\partial r} + 5 \right)^2 \left( M^0 + (1 - \Gamma) \langle \rho \rangle T - \tilde{X} \right) + \left( r \frac{\partial}{\partial r} + 5 \right) (\Gamma D_7 - D_5) \langle \rho \rangle \mu T \pm \frac{\partial \tilde{\Sigma}}{\partial t} \pm \langle \rho \rangle \frac{\partial}{\partial t} \hat{O}_3 Q
\end{equation}

\begin{equation}
\label{eq:equation1}
\pm \frac{\partial N}{\partial t} = \frac{\partial}{\partial r} \left( \widetilde{M} + \Gamma \langle P \rangle Q - \langle P \rangle Q + X \right) - (1 - \Gamma) \langle \rho \rangle \mu D_5 Q
\end{equation}

Now the above set of five equations describe higher order hydromagnetic turbulence.


\section{Example:}

The relations shown above soon gets difficult to track, as noticed by Chandrasekhar as well, in his work on hydromagnetics \cite{chandrasekhar:1955a, chandrasekhar:1955b}. However, the need of such work has always been undermined as energy spectra alone has been found to cater to the need to explain hydrodynamic as well as hydromagnetic turbulence well.

Below we discuss at least one example \cite{mukherjee:2019a} where, energy spectra alone has been found to fail to explain the numerical observation the authors performed.

A chaotic flow and uniform magnetic field is initialised. The initial condition is chosen such that, it is 3D MHD unstable. The flow and field variables are allowed to evolve numerically in a 3D periodic domain and the simulation is run enough for the system to get thermalised. Ideally, energy must have got equipartitioned once the system gets thermalised. However, the authors observe the following:

Even though, quite a number of the simulation results showed a thermalised energy spectrum \cite{mukherjee:2019b}, few have been found to reconstruct the initial flow as well as the field structures. In particular, an initial Taylor-Green flow with a uniform background magnetic field, is found to recur, while an initial Arnold-Beltrami-Childress flow with same uniform magnetic field, is found to thermalise.

The work was a first of its kind and has been now extended for several more initial flow and field variables \cite{mukherjee:2019c}. For example, an initial Robert's flow with a uniform background magnetic field, is found to recur, while an initial cat's-eye flow with same uniform magnetic field, is found to thermalise \cite{mukherjee:2019c}. Extension of such observation in terms of less restrictive conditions \cite{mukherjee:2018c, mukherjee:2019d, biswas:2021} have also been achieved. The authors have also reproduced the similar observations with different simulation packages, for example, finite-volume \cite{mignone:2007} as well as pseudo-spectral \cite{mukherjee:2018a, mukherjee:2018b, rupak_mukherjee_2021_4682188} algorithms for spatial discretisation and explicit and implicit time marching algorithms.

The general assumptions in all the works performed so far in regard to above nonlinear oscillations are the following: so far, `recurrence' phenomena have been observed for single-fluid plasmas, where, no large scale electric field is obtained due to charge separation between different species. In addition, it is assumed that a hydromagnetic description of plasma is good enough, such that, the individual fluid-elements can be assumed to posses a Maxwellian velocity distribution of the particles. This requires the plasma to be warm or cold, such that the plasma remains highly collisional within each fluid-element; such that any distortion of the parent Maxwellian distribution can get restored at a timescale faster than modes of any growing instability. Currently we are in pursuit of observing `recurrence' with two-fluid and fully kinetic models as well.


\subsection{Governing equations:}
A typical recurrence problem solves the following equations:\\
Continuity equation:
\begin{equation}
\label{continuity} \frac{\partial \rho}{\partial t} + \vec{\nabla} \cdot (\rho \vec{u}) = 0
\end{equation}
Momentum equation:
\begin{equation}
\label{momentum} \rho \frac{\partial \vec{u}}{\partial t} + \rho ( \vec{u} \cdot \vec{\nabla} ) \vec{u} = \vec{j} \times \vec{B} - \vec{\nabla} P + \mu {\nabla}^2 \vec{u} + \rho \vec{F}
\end{equation}
where the left hand side $\frac{D}{Dt} \equiv \frac{\partial}{\partial t} + \vec{u} \cdot \vec{\nabla} $ is the convective derivative, $\vec{j} \times \vec{B}$ is the Lorentz force acting on the fluid element, $\vec{\nabla} P$ represents the compressibility of the fluid and $\mu {\nabla}^2 \vec{u}$ is the viscous term with $\mu$ being the compressibilty of the fluid. Here in this simulation, we concentrate on decaying turbulence, with no external force acting on the fluid.\\
Equation of state:
\begin{equation}
\label{eos} \frac{d}{dt} \left(\frac{P}{\rho^\gamma}\right) = 0
\end{equation}
This follows that pressure is proportional to the density of the fluid.\\
Ampere's law:
\begin{equation}
\label{ampere} \vec{\nabla} \times \vec{B} = 4 \pi \vec{j}
\end{equation}
where we ignore the contribution from the displacement current being relativistically suppressed.
Faraday's law of induction
\begin{equation}
\label{faraday} \frac{\partial \vec{B}}{\partial t} = - c \vec{\nabla} \times \vec{E}
\end{equation}
Ohm's law for conducting fluid
\begin{equation}
\label{ohm} \vec{E} + (\vec{u} \times \vec{B} ) / c = \eta \vec{j}
\end{equation}
where, $\eta$ is the coefficient of resistivity.


\subsection{Initial conditions}
As mentioned in the earlier section, few of the initial conditions are found to recur. Two such examples are,\\

\begin{minipage}{0.45\textwidth}
\hspace{1cm} Taylor-Green Flow: 
\begin{eqnarray*}
u_x &=& A ~ U_0 \left[ \cos(kx) \sin(ky) \cos(kz) \right]\\
u_y &=& - A ~ U_0 \left[ \sin(kx) \cos(ky) \cos(kz) \right]\\
u_z &=& 0
\end{eqnarray*}
\end{minipage}
\begin{minipage}{0.45\textwidth}
\hspace{2cm} Roberts Flow:
\begin{eqnarray*}
u_x &=& A ~ U_0 \sin(kz)\\
u_y &=& B ~ U_0 \sin(kx)\\
u_z &=& C ~ U_0 \sin(ky)
\end{eqnarray*}
\end{minipage}\\

For completeness, we took two more initial flow structures, that have not been found to recur:\\

\begin{minipage}{0.45\textwidth}
\hspace{1cm} Arnold-Beltrami-Childress Flow
\begin{eqnarray*}
u_x &=& U_0 \left[ A \sin(k z) + C \cos (k y) \right]\\
u_y &=& U_0 \left[B \sin(k x) + A \cos (k z)\right]\\
u_z &=& U_0 \left[C \sin(k y) + B \cos (k x)\right]
\end{eqnarray*}
\end{minipage}
\begin{minipage}{0.45\textwidth}
\hspace{1cm} Cats-Eye Flow
\begin{eqnarray*}
u_x &=& U_0 ~B ~\sin(ky)\\
u_y &=& U_0 ~A ~\sin(kx)\\
u_z &=& U_0 \left[ A \cos(k x) - B \sin (k y) \right]
\end{eqnarray*}
\end{minipage}


\subsection{Simulation parameters}
We repeat the simulation reported earlier, and solve the single-fluid equations mentioned above (\ref{continuity}-\ref{ohm})  using different initial conditions, with the plasma parameters provided in the table below. The numerical simulation is performed using the simulation package TARA \cite{rupak_mukherjee_2021_4682188} on the Sikkim University `Brahmagupta' HPC system. We choose the following simulation parameters:\\

\begin{minipage}{0.45\textwidth}
\begin{center}
\begin{tabular}{||c | c | c | c||} 
 \hline
 $R_e$ & $R_m$ & $M_A$ & $M_s$ \\ [0.5ex] 
 \hline\hline
 $10^3$ & $10^3$ & $1$ & $0.1$ \\ [1ex] 
 \hline
\end{tabular}
\end{center}
\end{minipage}
\begin{minipage}{0.45\textwidth}

\begin{center}
\begin{tabular}{||c | c | c | c | c | c ||} 
 \hline
 $N$ & $L$ & $\rho_0$ & $U_0$ & $A=B=C$ & $k$ \\ [0.5ex] 
 \hline\hline
 $128$ & $2 \pi$ & $1$ & $0.1$ & $1$ & $1$ \\ [1ex] 
 \hline
\end{tabular}
\end{center}
\end{minipage}\\

As mentioned above, the results have been also reproduced at even higher resolution runs previously. In TARA, we use NVIDIA cuSolverRF with super-time-stepping algorithm that runs on our NVIDIA-A30 GPU card.


\subsection{Analysis}
As reported in all the recurrence related papers mentioned above, the energy spectra has been found to be inadequate in analysing the origin of such phenomena. This raises a fundamental question: is the description of MHD system via energy-spectrum alone, complete? We find that this is something that Chandrasekhar realised precisely, but has been overlooked for decades \cite{chandrasekhar:1956}. Thus, development of an alternate theory that gives a more realistic depiction of the system is beneficial for application purposes and understanding complex phenomena. Chandrasekhar also observed that such description of turbulence in terms of phase-relations becomes even more complicated for hydromagnetic systems \cite{chandrasekhar:1955a, chandrasekhar:1955b}. But this approach remains a largely unexplored territory to describe such phenomena displayed by the MHD system. In this work, we have tried to extend Chandra's work upto third-moment (of kinetic Vlasov equation) for hydro and magneto-hydrodynamic models. However, an appropriate theory is still under-developed. This paper doesn’t claim to solve the limitations inherent in Kolmogorov’s model; rather, it seeks to explore potential alternate frameworks that may offer deeper insights and represent the true behaviour of complex hydromagnetic or hydrodynamic fluids.


\section{Conclusion}
In summary, we venture to re-think turbulence in the way Chandrasekhar described it in 1950s. Our findings extend Chandrasekhar's deductive theory of turbulence to encompass higher order moments for hydrodynamics as well as hydromagnetics by deriving three governing equations from the moments of Vlasov equation. This framework incorporates correlation tensors to account for energy transfer between Fourier modes with different wave numbers. Through rigorous mathematical analysis, we bridge the gap between heuristic models and detailed phenomenological descriptions, enhancing our understanding of turbulence in electrically influenced environments. This advancement not only builds upon the foundational work of Chandrasekhar and Heisenberg but also provides a robust basis for future research in turbulence theory, offering new insights and predictive capabilities. \\

As a relevant example of the need of such analysis we recall a well established phenomena called `recurrence', where the energy spectra does not throw light to explain the effect. But by integrating a higher order model, we aim to be able to derive some useful insights on the complexities that traditional spectral methods might overlook. Our work, thus strengthens the motivation for exploring a more kinetic-based description of the dynamics of such a system.


\section*{Acknowledgement}
Both the authors are grateful to the referee(s) for their helpful comments. All the simulations reported in this manuscript are performed in Sikkim University's `Brahmagupta' HPC facility. One of the authors Rupak Mukherjee acknowledges IUCAA visiting associateship program for their kind support.

\bibliographystyle{unsrt}
\bibliography{biblio}
\end{document}